\documentclass{desyproc}

\usepackage{epsfig}
\usepackage{citesort}
\usepackage{graphicx}
\usepackage{amssymb}
\usepackage{color}

\begin{document}
%------------------------------------
\title{\vspace{-3cm}{\small \hfill{MPP-2010-136}}\\[1.8cm]
Hidden Photons from the Sun}

%for single authors the superscripts are optional
\author{{\slshape Davide~Cadamuro and Javier~Redondo}\\[1ex]
Max Planck-Institute f\"ur Physik, F\"ohringer Ring 6, D-80805 M\"unchen, Germany}

% if the proceedings are available online (e.g. at Indico)
% please enter the contribution ID or file_name below for the DOI
%\contribID{32}
\contribID{Redondo\_Javier}

% TO THE CONFERENCE EDITORS:
% please update the following information
% before sending the template to the authors
% \confID{800}  % if the conference is on Indico uncomment this line
\desyproc{DESY-PROC-2010-03}
\acronym{Patras 2010} % if you want the Acronym in the page footer uncomment this line
\doi  % if there is an online version we will register DOIs

%%%%%%%%%%%%%%%%%%%%%%%%%%%%%%%%%%%%%%%%%%%%

\maketitle

\begin{abstract}
A brief account of the phenomenon of photon oscillations into sub-eV mass hidden photons is given 
and used to estimate the flux and properties of these hypothetical particles from the Sun. 
A new generation of dedicated helioscopes, the Solar Hidden Photon Search (SHIPS) in the Hamburg Observatory
amongst them, will cover a vast region of parameter space.
\end{abstract}

%%%%%%%%%%%%%%%%%%%%%%%%%%%%%%%%%%%%%%%%%%%%

%\section{Motivation}

A Hidden Photon (HP) is the gauge boson of a local $U(1)$ hidden symmetry.
This kind of symmetries arise very frequently in many popular extensions of the Standard Model, especially in String Theory~\cite{Goodsell:2010ie}.
Known particles have no direct interaction with HPs (hence the name hidden) but couplings can still be generated through radiative corrections or gravity. In particular, HPs could be really not so hidden, because very massive particles with both electric and hidden charge can generate kinetic mixing with the standard photon~\cite{Holdom:1985ag},
\begin{equation}
\mathcal{L}_{mix}=-\frac{1}{4}\chi A_{\mu\nu}B^{\mu\nu}\; ,
\end{equation}
where $A_{\mu\nu}$ and $B_{\mu\nu}$ are the photon and HP field strengths. 
In this case, the natural value of the dimensionless kinetic mixing coupling $\chi$ is that of a quantum correction. 
Since the hidden gauge coupling can be very small, see e.g.~\cite{Goodsell:2009xc}, and there can be cancellations between different mediator contributions, there is no clear minimum for $\chi$. Values in the $ 10^{-16}\sim 10^{-3}$ range have been predicted in the literature~\cite{Goodsell:2009xc,Dienes:1996zr%,Abel:2008ai,Bullimore:2010aj}.
}.
Moreover, HPs can become massive via Higgs and/or Stueckelberg mechanisms~\cite{Ahlers:2008qc}.

This very weak interaction makes HPs perfect candidates for the dark sector that current cosmology and astrophysics are revealing. They have been proposed as Dark Matter (DM) candidates~\cite{Redondo:2008ec%,Pospelov:2008jk} 
}
and as mediating Dark Forces between DM particles~\cite{ArkaniHamed:2008qn}.
On the other hand, if their mass is in the meV range, their cosmological relic abundance could also provide the extra radiation favoured by the recent WMAP-7 results~\cite{Jaeckel:2008fi}.

%\section{Photon Oscillations}

The presence of the kinetic mixing term signals that the photon ($A_\mu$) and the HP ($B_\mu$) fields are not orthogonal. 
Since the photon is by definition an interaction eigenstate (couples to the electric charge) and the HP is in general massive, 
the kinetic mixing \emph{misaligns} the interaction ($A, S$) and propagation eigenstates ($\tilde{A}, B$). 
The photon can be written in terms of propagation eigenstates as
\begin{equation} \label{phot}
A_\mu=\tilde{A}_\mu-\sin \chi B_\mu\simeq\tilde{A}_\mu- \chi B_\mu\; .
\end{equation}   
Most importantly, the state orthogonal to it, i.e. 
\begin{equation}
S_\mu=B_\mu+ \chi \tilde{A}_\mu\; ,
\end{equation} 
is completely sterile to electromagnetic interactions.

The \textit{misalignment} between interaction and propagation eigenstates is well know to produce flavour oscillations, i.e. photons ($\gamma$) will convert into sterile states ($\gamma'$) --and vice-versa-- as they freely propagate with a probability~\cite{Okun:1982xi}
\begin{equation}
P(\gamma\rightarrow\gamma')=4\chi^2\sin^2\left(\frac{m^2_{\gamma'}L}{4\omega}\right) , 
\end{equation}
where $m_{\gamma'}$ is the HP mass, $L$ is the length of the path covered and $\omega$ is the photon energy. 

%\section{The Sun in Hidden Photons}

The oscillation mechanism previously described is presently the best known ally to search for $\sim$meV mass HPs by means of the so called Light Shining through Walls (LSW) experiments~\cite{Okun:1982xi,Jaeckel:2007ch%,Caspers:2009cj, Jaeckel:2008sz, Jaeckel:2009wm,Cameron:1993mr,Fouche:2008jk,Afanasev:2008fv,Ehret:2010mh}.
}. 
The experimental setup consists of a photon source and a photon detector separated by a thick shielding. Due to the very low probability of photon to HP conversion, very powerful sources, as high intensity lasers, and very sensitive photon detectors are necessary.

The most remarkable source of photons to be exploited could be our Sun itself. 
Most importantly, it can be an astounding source of hidden photons as well, if they are produced in $\gamma\to\gamma'$ oscillations of the photons populating its interior~\cite{Okun:1982xi,Popov:1991,Redondo:2008aa}. 
Solar Neutrino flux data constrain an exotic ``invisible" solar luminosity to be less than 
10\% of the solar photon luminosity~\cite{Gondolo:2008dd} ($L_\odot\simeq3.84\times 10^{26}$ W).
These HPs can be searched for with helioscopes such as CAST and SUMICO, built for the search of solar axions. Solar HPs can easily travel all along to the Earth, entering an optically thick vacuum vessel in which the inverse oscillation process $\gamma'\to \gamma$ will recreate solar photons that could in principle be easily detected.
A 1 m$^2$ Helioscope like CAST could detect the hidden photons emitted by the Sun inside a solid angle of about $(150\times 10^9)^{-2}$ sr, which means that the Sun could be providing up to 140 W of HPs without contradicting the solar dynamics.

Photon$\to$HP oscillations inside the Sun are strongly affected by matter effects. 
A complex refraction index $n$ is normally defined to account for the refractive and absorptive properties of a medium
\begin{equation}
-2\omega^2(n-1)\equiv m^2_\gamma+i \omega\Gamma \; .
\end{equation}
The optical properties of the solar plasma depend mainly on the 
electron density $n_e(R)$, which in the Sun is a decreasing function of the radius $R$, through the plasma frequency $\omega_p$ 
\begin{equation}\label{mgamm}
\omega_p^2(R)=m_\gamma^2(R)=\frac{4\pi\alpha}{m_e}n_e(R)\; .
\end{equation} 

Taking into account these considerations, it is possible to derive a very simple expression for the probability of a photon to convert into a HP inside the Sun and subsequently escape~\cite{Redondo:2008aa}
\begin{equation}
P(\gamma({\rm in})\rightarrow\gamma'({\rm out}))=\frac{m_{\gamma'}^4(R)}{\left(m_{\gamma}^2(R)-m_{\gamma'}^2\right)^2+\left(\omega\Gamma(R)\right)^2}\; .
\end{equation}
The dependence of $m_\gamma$ on $n_e$ given by \eqref{mgamm} makes the last equation quite interesting for our discovery purpose. 
The interior of the Sun has indeed a very high electron density ($n_e\simeq 10^{25}$ cm$^{-3}$) which corresponds to a rather large photon mass $m_\gamma\simeq 300$ eV. 
In this region the production of HPs of very small mass will be extremely suppressed.

Moving away from the solar centre the electron density decreases, allowing eventually the effective photon mass to be equal to HP mass 
at a certain $R_*$, i.e.~$m_\gamma(R_*)=m_{\gamma'}$. Since the Sun is a weakly coupled plasma (except for absorption lines and very small frequencies one has $\omega\Gamma\ll m_\gamma^2$), the resonant production of HPs in a thin spherical shell is very much enhanced and generally dominates over the emission from the rest of the Sun. 
In the sharp resonance approximation, i.e.~considering the HPs production only in this tiny shell whose thickness can be approximated by $\Delta R\simeq
\omega\Gamma(d m_\gamma^2/dR)^{-1}$, it is possible to linearise the Sun density profile
to obtain an analytical HP flux $\Phi_{\gamma'}$,
\begin{eqnarray}
\label{flux}
\frac{d\Phi_{\gamma'}}{d\omega}&=&\frac{1}{4\pi R_{earth}^2}\int^{R_\odot}_0 4\pi R^2dR\frac{1}{\pi^2}\frac{\omega\sqrt{\omega^2-m_{\gamma'}^2}}{e^{\omega/T}-1}\frac{\chi^2 m_{\gamma'}^4}{\left(m_{\gamma}^2-m_{\gamma'}^2\right)^2+\left(\omega\Gamma\right)^2}\Gamma\nonumber \\
&\simeq&\left.\frac{2}{\pi^2}\frac{R^2}{R_{earth}^2}\frac{\chi^2 m_{\gamma'}^4}{\omega\frac{dm_{\gamma}^2}{dR}}\frac{\omega\sqrt{\omega^2-m_{\gamma'}^2}}{e^{\omega/T}-1}\right|_{R=R_*}\; .
\end{eqnarray}
Interestingly, in this approximation the flux does not depend on the absorption coefficient.

From the Earth, HPs emitted from the resonant shell will appear most prominently in a ring of radius $R_*$ in the Sun. This could be in principle be used  to measure the HP mass through the estimate of $R_*$ and $m_\gamma$. 

In particular, for HPs in the sub-eV mass range, the resonance would happen in the outer layers of Sun and the radius $R_*$ becomes practically equal to 
the solar radius $R_\odot$, where the plasma is not fully ionized. Unfortunately, the presence of bounded electrons makes things more complicated and spoils the 
simple picture we have described. A gas with bounded electrons can be modelled as a superposition of damped oscillators, from which  the index of refraction can be written as
\begin{equation} \label{nightmare}
-2\omega^2(n-1)=m^2_\gamma+i \omega\Gamma=\frac{4\pi\alpha n_e}{m_e}\sum_j \frac{f_j}{\omega^2-\omega_j^2+i\omega\gamma_j}\; ,
\end{equation}
where $\omega_j$ is the frequency of the $j$-th oscillator, $f_j$ is the fraction of electrons that can be excited in that frequency and 
$\gamma_j={2\alpha\omega\omega_j}/({3 m_e})$ is the free-radiator damping constant. Of course, in a plasma like our Sun, corrections to this simple picture such as thermal and pressure broadening have in principle to be included. 
For light of frequency $\omega$ far from the absorption lines it is possible to split the sum over resonances into a IR ($\omega_k\ll\omega$) and a UV part ($\omega_k\gg\omega$),  
\begin{equation}
m^2_\gamma\simeq
\frac{4\pi\alpha n_e}{m_e}\left(\sum_{IR} f_j-\sum_{UV}f_j\left(\frac{\omega}{\omega_j}\right)^2\right)\; .
\end{equation}
The net effect of bounded electrons is to decrease $m^2_\gamma$ with respect to the naive fully-ionized plasma case. Therefore, the resonance displaces inwards the Sun with respect to the position it should have in the fully ionized model. Moreover, this displacement is frequency dependent so that the HPs emission ring is  somewhat broadened with a tendency of higher frequencies to be emitted from the ring's interior. 
The emission of HPs with frequencies corresponding to strong absorption lines will be widespread in the ring, first because $\Gamma$ is bigger and $\Delta R\propto \Gamma$ but also because on-resonance the contribution of the oscillator to $n$ is purely imaginary and the real part changes sign very steeply.

We are currently working on a detailed code to compute the hypothetical flux of HPs from the Sun taking into account the detailed properties of the index of refraction in the solar plasma. The resulting fluxes are crucial to interpret correctly the outcome of a number of on-going helioscope experiments looking for solar hidden photons such as possible parasitic experiments of the CAST and SUMICO axion searches and a dedicated experiment at the Hamburg Observatory, the solar hidden photon search (SHIPS). The order of magnitude derived from \eqref{flux} is however most promising. Everything points towards an improvement of more than one order of magnitude in sensitivity for $\chi$ with respect to the current CAST limit~\cite{Redondo:2008aa}, exploring a vast unknown territory in parameter space, which in particular contains the region which could explain the WMAP-7 preference for an extra radiation-like component of the universe~\cite{Jaeckel:2008fi}.

\end{document}